\documentclass [preprint, prd, aps, nofootinbib]{revtex4}

\begin{document}

\draft
\title{\large \bf A geometrical derivation of the Dirac equation}
\author{\bf Y. Jack Ng\footnote{E-mail: yjng@physics.unc.edu (Y.J. Ng)}
and H. van Dam}
\address{Institute of Field Physics, Department of Physics and
Astronomy,\\
University of North Carolina, Chapel Hill, NC 27599-3255\\}

\begin{abstract}

We give a geometrical derivation of the Dirac equation by considering a
spin-$\frac {1}{2}$ particle travelling with the
speed of light in a cubic spacetime lattice.  The
mass of the particle acts to flip the
multi-component wavefunction at the lattice sites.
Starting with a difference equation for the case of one spatial and one
time dimensions, we generalize the approach to higher dimensions.
Interactions with external electromagnetic and gravitational fields are
also considered.  One logical interpretation is that only at the lattice
sites is the spin-$\frac {1}{2}$ particle aware of its mass and the
presence of external fields.


\end{abstract}

\maketitle

\bigskip

\section{Introduction}

There are different ways to derive the Dirac equation.  But probably there
is no derivation more elegant than the one Dirac gave in his
book.\cite{Dirac}  The derivation based on Wigner's analysis of the
irreducible unitary representation of the ``Poincare group'' (the covering
group of the inhomogeneous proper othochronous Lorentz group) certainly is
also important.\cite{Wigner}  There is another intriguing derivation which
Feynman gave (for the $SO(1,1)$ case with one spatial dimension and one time
dimension) in his class\footnote{L. Brown, private communication (2001).}
and which was given as a problem in his
book with Hibbs\cite{Hibbs}.

In this paper we give another derivation of the Dirac equation.  Our
approach bears some resemblance to Feynman's and is based on Dirac's
observation that the instantaneous velocity operators of the
spin-$\frac {1}{2}$
particle (hereafter called by the generic name ``the electron'') have
eigenvalues $\pm c$ and that they anticommute.\footnote{The Hamiltonian
for a free electron is given by $H = c \alpha_{i} p_{i} +
\rho_{3}mc^2$ (in Dirac's notation) with anticommuting $\alpha_{i}$ and
$\rho_{3}$.  The $i$-th component of the velocity is $\frac {dx_i}{dt} =
(i\hbar)^{-1} [x_i , H] = c \alpha_i$.  Thus $\frac {dx_i}{dt}$
has as eigenvalues
$\pm c$ , corresponding to the eigenvalues $\pm 1$ of $\alpha_i$.  This
result is actually implied by the uncertainty principle.
Dirac\cite{Dirac} also shows that $c \alpha_i$ consists of two parts, a
constant part $c^2 p_i H^{-1}$, connected with the momentum by the
classical relativistic formula, and an oscillatory part whose frequency is
high, being at least $mc^2/  \pi \hbar$.}
(Hereafter, unless clarity demands otherwise, we set the speed of light
$c$, as well as Planck's constant $\hbar$, equal to unity.)
We assume spacetime to be
``filled'' with a four-dimensional cubic lattice with lattice length
$\Delta x = \Delta y = \Delta z = \Delta t = l$.  While it is natural to
take $l$ to be the Planck length ($\sim 10^{-33}$ cm), we will simply
take it to be a length very small compared to the electron's Compton
wavelength, the only intrinsic length available in the problem for a free
electron.  For the resulting difference equation, the zeroth order
term in $\Delta t$ gives a trivial identity while the first order term
yields the Dirac equation.

It is interesting that the Dirac equation is invariant under rotations and
Lorentz transformations, while the underlying spacetime lattice is not.
This situation is one which is not unfamiliar in the spatial dimensions
in condensed matter physics.  (But a related and
more intriguing result was that found by Snyder\cite{Snyder} more than half
a century ago, who showed that spacetime
being a continuum is not required by Lorentz invariance.)  In our approach,
an electron's propagation through spacetime can be visualized as
consisting of two steps:  the electron transfers from one spatial lattice site
to a neighboring site in one unit of time (thus travelling with the speed of
light) and at a lattice site the electron (its multi-component
wavefunction, to be more precise) is ``flipped'' by a mass
operator\footnote{This is in consonance with the mass term
being equivalent to a helicity flip.} and
interacts with external fields.

In the next section, we consider the case of one spatial dimension
(and one time
dimension).  We treat the case of higher spatial dimensions in Section III.
Interactions with external electromagnetic and gravitational fields are
considered in Section IV.  Further discussions are given in the last Section.

\section{$SO(1,1)$: (1 + 1)-dimensional case}

We assume that the electron of mass $m$ moves with the speed of light from
one lattice
site to a neighboring (spatially left or right) site with time $t$ always
increasing on the ``cubic''
spacetime (z,t) lattice.  The wavefunction has two components
\begin{equation}
\psi (z,t) = \left( \begin{array}{c}
                     \psi_{+}(z,t) \\ \psi_{-}(z,t)
                     \end{array}  \right),
\label{psi1}
\end{equation}
where $\psi_{+}$ denotes the component arriving from the event $(z - \Delta t,
t - \Delta t)$ while $\psi_{-}$ means arriving from $(z + \Delta t, t - \Delta
t)$.

Next we assume that, at the lattice site $(z, t)$, the arriving components
are partially turned around by a unitary matrix:
\begin{equation}
\left( \begin{array}{c}
           \psi_{+} (z, t) \\ \psi_{-} (z, t)
          \end{array}   \right)
  = \mathcal{F}  \left( \begin{array}{c}
                              \psi_{+} (z - \Delta t, t - \Delta t) \\
                              \psi_{-} (z + \Delta t, t - \Delta t)
                             \end{array}    \right),
\label{disDir}
\end{equation}
with the ``flip operator'' $\mathcal{F}$ defined by
\begin{equation}
\mathcal{F} \equiv e^{-iFm \Delta t}.
\label{flip}
\end{equation}
Here $F$ is a hermitian $2 \times 2$ matrix which we call the ``flip
matrix'' and give the most obvious form
\begin{equation}
F = \sigma_{x} \equiv \sigma_{1} = \left( \begin{array}{cc}
                                         0 & 1 \\
                                         1 & 0
                                        \end{array}  \right),
\label{sigma1}
\end{equation}
with $\sigma_{1}$ being the first Pauli matrix.  We will approximate Eq.
(\ref{disDir}) by a differential equation by first writing
\begin{equation}
\left( \begin{array}{c}
           \psi_{+}(z - \Delta t, t - \Delta t) \\
           \psi_{-}(z + \Delta t, t - \Delta t)
          \end{array}   \right)
        = \mathcal{T} \psi(z, t),
\label{jump1}
\end{equation}
with the ``transfer'' operator $\mathcal{T}$ given by
\begin{equation}
\mathcal{T} = e^{- \Delta t \left(\frac {\partial}{\partial t} +\sigma_3
\frac
{\partial}{\partial z}\right)},
\label{jump2}
\end{equation}
where $\sigma_3 \equiv diag (1, -1)$ is the third Pauli matrix.
Then the difference equation Eq. (\ref{disDir}) takes the form
\begin{equation}
\psi (z, t) = \mathcal{F} \mathcal{T} \psi (z, t).
\label{allDir}
\end{equation}

The difference equation becomes a differential equation if we limit
ourselves to the zeroth order (given by the identity
$\psi (z, t) = \psi (z,t)$) and the first order term in $\Delta t$.
The first order equation is
\begin{equation}
i \frac {\partial}{\partial t} \psi (z, t) = m \sigma_1 \psi (z, t)
-i \sigma_3 \frac {\partial}{\partial z} \psi (z, t),
\label{Dirac}
\end{equation}
the Dirac equation\footnote{In our representation of the Dirac matrices,
the positive-energy spinor for
the plane-wave solution takes the form, aside from a normalization
constant, $u(p) \sim (-\sigma_1 \sqrt{p^2 + m^2} - i \sigma_2 p - m)u(0)$,
with $u(0)$ being the two-component spinor $(1, -1)$, and the
negative-energy spinor $v(p) \sim (\sigma_1 \sqrt{p^2 + m^2} + i \sigma_2 p
- m)v(0)$, with $v(0) = (1, 1)$.} in $(1 + 1)$ dimensions!  There is no spin
in the
$SO(1,1)$ case, the little group of $p^{\mu}$ being trivial.
In passing we mention that,
for the Dirac equation to hold to all orders in $\Delta t$, due to the
fact that $\sigma_1$ and $\sigma_3$ do not commute with each other, it is
necessary to replace $\mathcal{F} \mathcal{T}$ in Eq. (\ref{allDir}) by
their ``symmetrical'' product $e^{-\Delta t (im \sigma_1 + \frac
{\partial}{\partial t} + \sigma_3 \frac {\partial}{\partial z})}$.  We will
not pursue this issue any further and will assume that $\Delta t$ is
sufficiently small that higher order terms are negligible, i.e.,
the Dirac equation is a good approximation to the original difference
equation.

\section{Higher spatial dimensions}

We start by reminding ourselves that, for the Dirac equation, the velocity
operators $\frac {1}{i \hbar} [ \vec{x}, H ] = c \vec{\alpha}$ not only
have eigenvalues $\pm c$, but they also anticommute with
each other.\footnote{Here an analogy with spin can be made.  Just as one
cannot specify all three
components of the spin simultaneously in quantum mechanics without running
into inconsistencies (of predicting a spin $\sqrt{3}$ times bigger in
some directions),
one cannot
specify all three components of the instantaneous velocity simultaneously
without running into inconsistencies of predicting a tachyonic speed of
$\sqrt{3}$ times the speed of light.}  The
latter fact makes the generalization of our approach to more than one
spatial dimension non-trivial.  Before we proceed to the $(3 +
1)$-dimensional case, let us first discuss the two spatial dimensional
$SO(2, 1)$ case.

As in the preceeding section (for the $SO(1, 1)$ case), we use $\sigma_3$
to give the dependence on the $z$ coordinate and $\sigma_1$ for the
flip
operator.  Of the three Pauli matrices, we have only $\sigma_2$ left; so
let us call the second spatial coordinate the $y$ coordinate.  Now the
problem is to express the dependence of $\psi$ on $y$ in the sense that
$\sigma_3$ gives the dependence on $z$.

To solve this problem we appeal to rotational invariance and make explicit
use of the spin-$\frac {1}{2}$ property of $\psi$.  Let $\mathcal{R}$ be the
rotation over $\pi /2$ from the $y$ axis to the $z$ axis.  Then
$\mathcal{U} (\mathcal{R})$, which represents this rotation, is given by
$\mathcal{U} (\mathcal{R}) = \frac {1}{\sqrt{2}} (1 - i \sigma_1)$.  Thus,
to the $\sigma_3 \frac {\partial}{\partial z}$ term in the ``transfer''
operator $\mathcal{T}$ we add
\begin{equation}
\mathcal{U}^{-1} (\mathcal{R}) \sigma_3 \frac {\partial}{\partial y}
\mathcal{U} (\mathcal{R}) = \sigma_2 \frac {\partial}{\partial y},
\label{rot}
\end{equation}
and obtain
\begin{equation}
i \frac {\partial}{\partial t} \psi (y, z, t) = \left( m \sigma_1
-i \sigma_2 \frac {\partial}{\partial y}
-i \sigma_3 \frac {\partial}{\partial z} \right) \psi (y, z, t).
\label{Dirac21}
\end{equation}
Note that the flip operator, which is used in the preceeding section to
invert the $z$-motion, also inverts the $y$-motion.  In Dirac's
notation\cite{Dirac}, we identify $\sigma_1 = \alpha_m$,
$\sigma_2 = \alpha_2$, $\sigma_3 = \alpha_3$.

Now we recall the general rule\cite{wein1} that spinors in $2n$ dimensions
and in $2n + 1$ dimensions have $2^n$ components.  Thus for the case of
three spatial dimensions and one temporal dimension ($SO(3, 1)$), we need
$4$-spinors.  And we need, besides $\alpha_m$, $\alpha_2$, $\alpha_3$, an extra
$\alpha$, all four of which anticommute with one another and have
eigenvalues $\pm 1$.  Following Dirac\cite{Dirac}, we introduce two
independent sets of Pauli matrices $\vec{\rho}$ and $\vec{\sigma}$.  The
$\rho$'s and the $\sigma$'s anticommute among each set, whereas the
$\rho$'s and the $\sigma$'s commute.  As we want to make $\alpha_m$ look
like a flip matrix, we pick
\begin{equation}
\alpha_m = \rho_1 \sigma_1 = \left( \begin{array}{cccc}
                                        0 & 0 & 0 & 1 \\
                                        0 & 0 & 1 & 0 \\
                                        0 & 1 & 0 & 0 \\
                                        1 & 0 & 0 & 0
                                       \end{array}   \right).
\label{alpham}
\end{equation}
We complete the set (by the same argument we have used above for the
$SO(2, 1)$ case) with
\begin{equation}
\alpha_1 = \rho_2 \sigma_1,\hspace{.2in} \alpha_2 = \rho_3 \sigma_1,
\hspace{.2in} \alpha_3 =
\sigma_3 {\bf 1}.
\label{alpha123}
\end{equation}
Here {\bf 1} is the $2 \times 2$ unit matrix.
In passing, we mention that it is easy to treat the case of $(4 +
1)$-dimensional spacetime, as we can now identify $\alpha_4$ with
$\sigma_2 {\bf 1}$.

For the case of $(3 + 1)$ dimensions, the equation we obtain is
\begin{equation}
i \frac {\partial}{\partial t} \psi = \left( \alpha_m m
- \alpha_1 i \frac {\partial}{\partial x}
- \alpha_2 i \frac {\partial}{\partial y}
- \alpha_3 i \frac {\partial}{\partial z} \right) \psi.
\label{Dirac31}
\end{equation}
It is of relevance to remark that the last three terms on the right hand
side of Eq. (\ref{Dirac31}) approximate the small finite steps of motion
with the speed of light before the event $(x, y, z, t)$ is reached.
The term $\alpha_m m$ represents the unitary transformation $e^{-iFm dt}$
which takes place at that event (see Eqs. (\ref{disDir}) and
(\ref{flip})).  Thus the electron is not aware of the fact that it has a
mass until it hits a lattice site.  If it has no mass, then it is not
flipped and it moves at the constant speed of light.  If it is massive and
is at ``rest'', then it must be that the electron zigzags around with
the speed
of light and returns to
its original spatial lattice site and wanders around again and returns
again etc.

\section{Interactions with external fields}

To put the Dirac equation in a covariant form, we follow the usual
procedure of writing $\alpha_m = \gamma^0$ (with $(\gamma^0)^2 = 1$) and
multiplying Eq. (\ref{Dirac31}) by $\gamma^0$ to yield
\begin{equation}
\left( i \gamma^{\mu} \partial_{\mu} - m \right) \psi = 0,
\label{covar}
\end{equation}
where ($\mu$ running over $0, 1, 2, 3$)
\begin{equation}
\gamma^{\mu} \equiv \gamma^{0} \alpha_{\mu},
\label{gammamu}
\end{equation}
with $\alpha_0 \equiv I$, the identity matrix and $\partial_{\mu} \equiv
\left( \frac{\partial}{\partial t}, \frac {\partial}{\partial x},
\frac {\partial}{\partial y}, \frac {\partial}{\partial z} \right)$.
From the way we have derived the Dirac equation, we can trace the $i
\gamma^{\mu} \partial_{\mu}$ term to the ``transfer'' of the electron at the
speed of light between the lattice sites while the $m$ term comes from the
``flip'' unitary transformation at the lattice sites.

The introduction of an electromagnetic field $A_{\mu}$ is straightforward
by using
the prescription of minimal substitution in Eq. (\ref{covar})
\begin{equation}
i \partial_{\mu} \Rightarrow i \partial_{\mu} + e A_{\mu}.
\label{em}
\end{equation}
Although the $e \alpha_{\mu} A_{\mu}$ term goes together with the
$\alpha_{\mu} \partial_{\mu}$ term in the minimal subsitution rule, it
is tempting to keep the $\partial_{\mu}$ term identified with the transfer
between lattice sites and put the $e \alpha_{\mu} A_{\mu}$ term together
with
the $Fm$ flip term as taking place at the lattice sites.  (But we should
keep in mind that, since the Dirac matrices do not commute among
themselves, beyond the first order term, there is a difference between
associating the interaction term with the ``transfer'' operator
$\mathcal{T}$ or the ``flip'' operator $\mathcal{F}$.\footnote{For the
$SO(1, 1)$ case, the $A_{\mu}$ term can be incorporated into
the flip operator in Eq. (\ref{disDir}).})

To incorporate gravitational interactions one needs the tetrad (or
vierbein) formalism\cite{UandK}.  One introduces at every event $x$ a set
 of local inertial coordinates with a tetrad
$e_a^{\mu}$ of axes labelled by the
Minkowski index ${a, b, c}$ running over $0, 1, 2, 3$.\cite{Schwinger}
Then the metric in any general noninertial coordinate system is given by
$g_{\mu \nu} = e_{\mu}^a \eta_{ab} e_{\nu}^b$ in terms of the flat
Minkowski metric $\eta_{ab}$.  Gravitational interactions are introduced
via
the substitution rule
\begin{equation}
\gamma^a \partial_a \Rightarrow \gamma^a e_a^{\nu} (\partial_{\nu}
 - \frac {1}{4} i \omega_{bc \nu} \sigma^{bc} ),
\label{covder}
\end{equation}
where $\omega^b_{a \nu} = (\partial_{\nu} e_a^{\mu} + \Gamma_{\nu
\lambda}^{\mu} e_a^{\lambda}) e^b_{\mu}$ in terms of the affine connection
$\Gamma_{\nu \lambda}^{\mu}$ and $\sigma^{bc} = \frac {1}{2}i[ \gamma^b,
\gamma^c]$.  At every spacetime lattice site labelled by $x$, we have a
tetrad $e_a^{\mu}$.  In our interpretation, the electron travels with the
speed of light between lattice sites; this is represented by
$\gamma^a e _a^{\nu} \partial_{\nu}$.  Then at the lattice site
there is a unitary transformation
which, in addition to the mass ``flip'', now contains the interaction term
$\gamma^a e_a^{\nu} \omega_{bc \nu} \sigma^{bc}$.

\section{Discussions}

We have presented a novel derivation of the Dirac equation, hoping to
shed new light on the physics of the electron.  Motivated
by the distinct possibility that the underlying spacetime is discrete at
small scales, we have started with a discrete ``cubic'' lattice.  The
resulting Dirac equation emerges as the lowest nontrivial order of
approximation.  Thus the observed Lorentz invariance does not preclude the
existence of a discrete spacetime at small scales.

Is our approach useful?  We think so.
(1) The very fact that the underlying spacetime is discrete
means that there is automatically an ultraviolet cutoff which may be used
to ameliorate divergence problems in nonrenormalizable theories like
(perhaps) quantum gravity.  (2) Our starting point is a difference equation
rather than a differential equation.  While difference equations are more
tedious to deal with analytically, they may hold some advantages in
numerical calculations.

We conclude with some speculations and a couple of open questions.
In the scenario we have proposed, the electron travels between lattice sites
with the speed of light.  Only at the lattice sites does the electron
``feel'' its mass and perhaps also the presence of all external
fields.\footnote{It is natural to visualize interactions taking place at
lattice sites
(in conjunction with the mass flip).  After all, interactions occur at
spacetime points and spacetime points are the lattice sites if the
underlying spacetime is discrete.}
(Since it is a Yukawa-type interaction which, via the Higgs
mechanism, generates mass for the electron, it seems reasonable to
assume that at least Yukawa-type interactions take place only at the
lattice sites where the mass operator makes its presence felt.)  But if
gravitational
interactions also take place mainly at the lattice sites, does that mean
spacetime vertices somehow play an important role in
concentrating curvature?  And if so, how is this description of geometry and
topology related to the
Regge calculus\cite{Regge}, for example?  These problems deserve further
investigations.\\

\bigskip

\section*{Acknowledgements}

We thank Giovanni Amelino-Camelia and E. Merzbacher for useful
discussions, Laurie Brown for a useful correspondence, and
C.N. Yang for his encouragement to understand the Dirac equation from
a new perspective.  We also thank Ted Jacobson and A. Rivero for calling
our attention to some useful references.
This work was supported in part by DOE under
\#DE-FG05-85ER-40219 and by the Bahnson Fund of University of North
Carolina at Chapel Hill.

\end{document}